\documentclass{article}
\usepackage{spconf,amsmath,graphicx}

\usepackage{enumitem}
\usepackage{siunitx}
\setlist{nosep, leftmargin=14pt}

\usepackage{mwe} 
\usepackage{xcolor}

\usepackage[hidelinks]{hyperref}

\newcommand{\Lagr}{\mathcal{L}}

\usepackage{multirow}

\title{An augmentation strategy to mimic multi-scanner variability in MRI \sthanks{This paper has been accepted for presentation at the International Symposium on Biomedical Imaging (ISBI 2021). \textcopyright 2021 IEEE.  Personal use of this material is permitted.  Permission from IEEE must be obtained for all other uses, in any current or future media, including reprinting/republishing this material for advertising or promotional purposes, creating new collective works, for resale or redistribution to servers or lists, or reuse of any copyrighted component of this work in other works.}}
\name{\begin{tabular}{@{}c@{}}
      Maria Ines Meyer$^{\dagger \star }$ \quad
      Ezequiel de la Rosa $^{\ddagger \star}$ \quad
      Nuno Barros $^{\star}$ \quad \\
      Roberto Paolella $^{\circ \star}$ \quad 
      Koen Van Leemput $^{\dagger \diamond}$ \quad
      Diana M. Sima $^{\star}$
      \end{tabular}}

\address{$^{\dagger}$ Health Technology Dpt., Technical University of Denmark, Lyngby, Denmark \\
         $^{\star}$ icometrix, Leuven, Belgium  \\
         $^{\ddagger}$ Dpt. of Computer Science, Technical University of Munich, Munich, Germany\\
         $^{\circ}$ imec Vision Lab, University of Antwerp, Antwerp, Belgium \\
         $^{\diamond}$ Martinos Center for Biomedical Imaging, MGH $\&$ Harvard Medical School, Boston, MA, USA 
         }
         
\begin{document}
%
\maketitle
\begin{abstract}
Most publicly available brain MRI datasets are very homogeneous in terms of scanner and protocols, and it is difficult for models that learn from such data to generalize to multi-center and multi-scanner data. We propose a novel data augmentation approach with the aim of approximating the variability in terms of intensities and contrasts present in real world clinical data. We use a Gaussian Mixture Model based approach to change tissue intensities individually, producing new contrasts while preserving anatomical information. We train a deep learning model on a single scanner dataset and evaluate it on a multi-center and multi-scanner dataset. The proposed approach improves the generalization capability of the model to other scanners not present in the training data.


\end{abstract}

\section{Introduction}
\label{sec:intro}

Magnetic resonance imaging (MRI) is regularly used for neuroimaging both in research and clinical settings. In this context convolutional neural networks (CNN) have been applied to several problems, such as the segmentation of different brain structures. However, these algorithms remain sensitive to factors such as hardware and acquisition settings, which can be especially problematic when integrating data from different cohorts \cite{Martensson2020}.
For these models to generalize to data collected using new or unseen scanners, large multi-center and multi-scanner datasets are necessary when training. Nevertheless, collecting such data is not trivial and most publicly available datasets are homogeneous in terms of scanner types and acquisition protocols. As a result, data availability has become a significant obstacle which hinders the application of deep learning-based models in clinical settings.

A widespread approach to deal with this type of problem is data augmentation (DA). The idea behind DA is simple: by applying transformations to the labeled data it is possible to artificially increase the training set, which implicitly regularizes the network. 
Basic operations such as geometric transformations, noise injection and random erasing can be found in a wide variety of applications, as detailed in \cite{Shorten2019}. 

In the medical imaging field DA is especially important. Although the aforementioned transformations can alleviate overfitting, they do not take into account the high variability in contrast found in some modalities. Some works have attempted to overcome this limitation by generating completely synthetic images using generative adversarial networks, as in \cite{Shin2018}. Others explore the color and contrast characteristics of the existing images, as is the case in \cite{Galdran2017}, where the training images were transformed to simulate different illumination conditions at acquisition. This type of approach was also explored for brain imaging. A scheme that generates synthetic versions of the training images such that they appear to have been acquired using different sequences is described in \cite{Jog2019}. This method has the disadvantage that it requires nuclear magnetic resonance parameter maps of the training images, which are often unavailable. 
Recently, Billot et al.~\cite{billot2020learning} proposed a generative model conditioned on a segmentation map to produce synthetic images on-the-fly with random appearance, deformation, noise, and bias field, improving the performance of a CNN by exposing it to often unrealistic contrasts. 

In the present work we propose a novel DA approach with the aim of reducing the scanner bias of models trained on single-scanner data. 
We randomly modify the MRI tissue intensities with a structural information preserving Gaussian Mixture Model based approach.
As a result the contrast between tissues varies, as seen when different scanners are used during acquisition. The proposed method does not depend on previous segmentations, is simple, fast and can be used on-the-fly during training. We illustrate the method in the context of brain substructure segmentation and evaluate the performance on multi-scanner patient data. We observe a clear improvement in generalization to unseen scanner types when adding the proposed method to the training pipeline.

\section{GMM-based intensity transformation}
\label{subsec:gmm_da} 

The idea behind the proposed approach is to increase the intensity and contrast variability of a single-scanner dataset such that it is representative of the variability found in a large multi-scanner cohort. 
It is well documented that in T1w brain MRI characteristic peaks in the histogram correspond to different tissues, \textit{i.e.,} CSF has the lowest intensity, followed by GM and WM. This has been explored by segmentation methods based on Gaussian Mixture Models (GMM).

GMM is a type of probabilistic model that assumes that data can be modeled as a superposition of $K$ Gaussians.  
Within this framework we can describe the intensities of each voxel $v$ in an image $I$ as: $p(v) = \sum_{k=1}^{K} \pi_k \mathcal{N}(v | \mu_k, \sigma^2_k)$. Each $\mathcal{N}(\mu_k, \sigma^2_k)$ is a \textit{component} of the mixture, with its own mean $\mu_k$ and variance $\sigma^2_k$, and $\pi_k$ are the mixing coefficients. 
For healthy or lesion-free brains, a 3-component GMM  is a good approximation of the tissue intensities in T1w brain images (after skull stripping), corresponding to the CSF, GM, and WM classes. 
After we estimate these components, we can predict the probability of each class label $C$, $p(C|v)$, and access the individual $\mu_k$ and $\sigma^2_k$. 
Thus we can modify images in the training data by changing their GMM probability distributions while preserving the inherent image characteristics.

We start by estimating the range of typical variation for each component from a large multi-scanner collection of patient data (dataset III in section \ref{subsec:datasets}). To do this, all images are first skull stripped, intensities are clipped at percentiles 1 and 99 to exclude noise, and normalized to the range $[0, 1]$. Then we fit a 3-component GMM to each image in the dataset. Mean and variance values for the 3 components in our data have approximate values and  standard deviations of $\mu= \{1, 2, 3\}\mathrm{\times 10^{-1}}, \quad s(\mu)=\{3,6,8\}\mathrm{\times 10^{-2}}$ and $\sigma^2= \{2, 1, 1\}\mathrm{\times 10^{-3}}, \quad s(\sigma^2)= \{1, 1, 3\}\mathrm{\times 10^{-3}}$, respectively.
Now that we have these values we generate new parameters $\mu \rightarrow \mu^\prime$ and $\sigma^2 \rightarrow \sigma^{2\prime}$ for the $k$-th component in an individual skull stripped image by:
\textbf{a)} sampling variation terms $q_{\mu}$ and $q_{\sigma^2}$ for each component from
the uniform distribution $U(-s(\mu)_k, s(\mu)_k)$ and $U(-s(\sigma^2)_k, s(\sigma^2)_k)$, respectively, 
and \textbf{b)} adding these values to the original parameters, such that  $\mu^\prime_{k}=\mu_k + q_{\mu}$ and $\sigma^{2\prime}_{k}=\sigma^2_k+q_{\sigma^2}$. Essentially we are creating a new intensity distribution for each of the tissues.
The choice of a uniform distribution for sampling the new variation terms implies that any random combination of tissue intensities can be generated. We could restrict this to more probable distributions by selecting a normal distribution. However, since exposing networks to unrealistic augmentation is beneficial for learning \cite{billot2020learning}, we decided to allow the possibility for some unrealistic combinations to arise.

\begin{figure}[t!]
    \centering
    \includegraphics[width=0.45\textwidth]{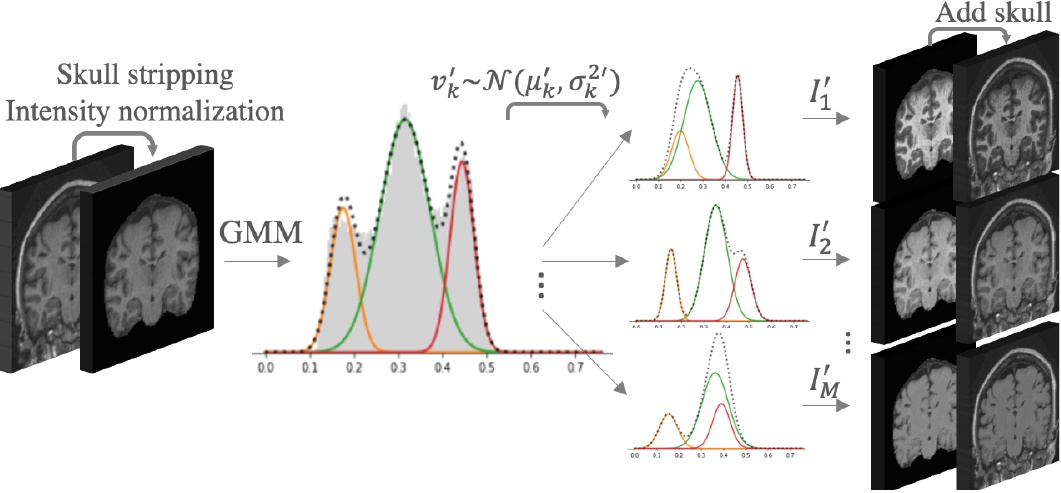}
    \caption{Diagram of the main steps in the proposed method.} 
    \label{fig:gmm-augmentation}
\end{figure}

Once the new components have been defined, we could use common histogram matching to generate the new images. However, this does not guarantee that structural information is preserved (e.g.,  two components could overlap or even shift order, and voxels from one tissue would be wrongly assigned to another class).
To avoid this we describe the intensity $v$ of some voxel in terms of the distance from the mean of the component: we compute the Mahalanobis distance $d_{vk} = (v-\mu_k)/\sigma_k$.
This implies that if we know the values of $\mu_k$ and $\sigma^2_k$ we can find the updated value of $v \rightarrow v^\prime$ for each component $k$:
$v^\prime_k = \mu^\prime_k + d_{vk} \sigma^{\prime}_k $. 
Fig. \ref{fig:gmm-augmentation} shows a depiction of the method. An implementation will be available at \url{https://github.com/icometrix/gmm-augmentation}

\subsection{Available Datasets}
\label{subsec:datasets}
\textbf{I) OASIS-1}
Contains T1w MRI scans from 416 subjects (age: $[18, 96]$). Only the $316$ healthy subjects were considered. The data was randomly split into train/validation/test sets ($n=179/45/56$). All images were acquired on a 1.5T Siemens Vision scanner, using MP-RAGE protocol \cite{Marcus2007}. 

\noindent\textbf{II) MICCAI 2012} Comprises 35 T1w healthy images from the OASIS dataset with manual labels for use in the MICCAI 2012 Grand Challenge and Workshop on Multi-Atlas Labeling \cite{MICCAI2012}. All images were removed from the OASIS dataset, to avoid overlap. We exclude 5 repeated subjects and use the remaining for evaluating the methods on the manual labels. 

\noindent \textbf{III) MS Dataset} Collection of multi-center T1w MRI scans from 421 individual Multiple Sclerosis (MS) patients. Very heterogeneous, $e.g.$ age ($[16, 81]$y), sex (M/F $\sim 33\%$/$67\%$), slice thickness in T1 ($[0.4, 1.5]$ mm), magnetic field strength ($1.5$T/$3$T $\sim 43\%$/$57\%$), scanner manufacturer (Philips, GE, Siemens and Hitachi) and model ($29$ devices). We sampled a set of $89$ images from $10$ different scanner models to use as test set. For an additional experiment we pooled a train/validation set of $180/49$ images (to preserve a similar size to OASIS), making sure not to include any scanner models present in the test set or in OASIS. The complete dataset was used to estimate the range of typical variation for the different tissues, as described in the previous section.
  
 Due to scarcity of manual delineations, we use brain substructure delineations obtained with ico\textbf{brain} \textbf{ms}, a clinically available and FDA-approved software \cite{Jain2015}, for training and evaluating the models.
All images were normalized using a modified z-score function robust against outliers, where the median of the distribution was preferred over of the mean, and the standard deviation of the distribution was computed within percentiles $10$ and $90$.
Additionally, images were bias-field corrected using the  N4 inhomogeneity correction algorithm as implemented in the Advanced Normalization Tools (ANTs) toolkit \cite{Tustison2010_n4} and linearly registered to MNI space  using the tools implemented in NiftyReg \cite{Ourselin2001_niftyreg}. 

\section{Experiments and Results}
\label{sec:results}
\paragraph*{Experimental setup}{
We investigate the added value of the described method to the task of brain structure segmentation. We trained a CNN to segment white matter (WM), gray matter (GM), cerebro-spinal fluid (CSF), lateral ventricles (LV), thalamus (Tha), hippocampus (HC), Caudate Nucleus (CdN), Putamen (Pu) and Globus Palidus (GP).
We compare three different models: \textit{Base} is trained on OASIS; \textit{BaseDA} is trained on OASIS with the addition of the DA strategy; \textit{BaseMS} is trained on the MS Dataset (no DA). 
}
\paragraph*{Model Architecture}{ 
We use a conventional 3D UNet architecture \cite{Cicek2016} with some changes, namely: weight normalization layers \cite{Salimans2016} are added after each convolutional operation instead of batch normalization; LeakyReLU is used as the main activation function and strided convolutions are used instead of max pooling. 
The network takes as input patches of size $128\times128\times128$ and outputs probability maps of size $88\times88\times88$. As loss function we use a combination of the  weighted categorical cross-entropy loss ($\Lagr_{\textrm{w}CE}$) with a Dice loss ($\Lagr_{Dice}$), such that $\Lagr  = \Lagr_{\textrm{w}CE}  + \Lagr_{Dice}$ \cite{Isensee2019}. Individual class-specific weights are estimated on the training set. 
Kernel size is $3\times3\times3$ and initial number of filters 16 (raised to the power of $2$ at each layer in the encoder path). All hyperparameters are tuned on a subset of the training and validation sets. 
The model is trained until convergence using mini-batch stochastic gradient descent (Adam optimizer) with initial learning rate $\lambda=0.001$ on a machine equipped with a Tesla K80 Nvidia GPU (12 GB dedicated). 
}
\paragraph*{Performance metrics}{ 
Dice scores ($DC$), sensitivity ($Se$), precision ($Pr$) and percentage of outliers ($\% outliers$) are reported.   Here any point that is farther than $1.5IQR$ (interquartile range) from the upper or lower quartiles of the distribution is considered an outlier. Recall that $Se$ and $Pr$ measure the presence of false negatives and false positives, respectively. $DC$ values are compared using Wilcoxon paired rank-sum and Levene tests to evaluate the null hypotheses $H_0$ that the results from the different models have equal median and variance values, respectively. These tests were selected given the presence of outliers and deviations from normality in the distributions, as seen in Fig. \ref{fig:boxplots_results}. Results are summarized in terms of median ($P50$) and percentile 10 ($P10$).  
                                }
\paragraph*{Results and Discussion}{\mbox{}\\
\noindent\textbf{OASIS}
The models achieve high Dice scores and low variability with overall small incidence of outliers (Table \ref{table:oasis-test}). $Se$ and $Pr$ are very similar for \textit{Base} and  \textit{BaseDA} (min: $Se_{GM}=0.87, Pr_{GM}=0.87;$  mean: $ \bar{Se}=0.94, \bar{Pr}=0.94$) and slightly lower for \textit{BaseMS} (min: $Se_{GM}=0.83, Pr_{GM}=0.88; $  mean:  $  \bar{Se}=0.92, \bar{Pr}=0.92$). 
For the \textit{Base} and \textit{BaseDA} models there is no statistical difference between the results (Wilcoxon: $p>0.05$, Levene: $p>0.05$), except for WM and GM, where \textit{Base} tends to perform better (Wilcoxon, $p<0.05$).
\textit{BaseMS} tends to underperform, with statistical differences in median  and variance (Levene: $p<0.05$ for WM, Tha, HC and Pu) (see Fig. \ref{fig:boxplots_results}). 
For datasets where the same scanners and sequences are present in the training and test sets there is no or only a minimal drop in performance when adding DA. We expect \textit{BaseMS} to perform worse, since the MS Dataset does not contain images with the same characteristics as OASIS.

\begin{table}[t]
\centering
\scriptsize
\caption{Summary of segmentation results on the OASIS test set -- automated segmentations. $DC$: Dice scores. Best results are highlighted.}
\setlength{\tabcolsep}{2.2pt} 
\renewcommand{\arraystretch}{1.1} 
\begin{tabular}{l|c|cccccccc|c }
\hline
Model                               &	Metrics     &	\multicolumn{8}{l}{OASIS test set}  \\

        &       &WM   &	GM  & LV   &	Thal &	HC   &	CdN & Pu & GP & All \\
\hline\hline
\multirow{2}{3.1em}{\textit{Base}}     &	$DC$ (P50)    &	$0.94$ & $\mathbf{0.90}$ & $\mathbf{0.96}$ & $0.95$ & $\mathbf{0.91}$ & $\mathbf{0.93}$ & $\mathbf{0.93}$ & $\mathbf{0.91}$ & $\mathbf{0.93}$ \\
                                     &	$DC$ (P10)    &	$\mathbf{0.93}$ & $\mathbf{0.87}$ & $\mathbf{0.94}$ & $\mathbf{0.94}$ & $\mathbf{0.90}$ & $\mathbf{0.92}$ & $\mathbf{0.92}$ & $\mathbf{0.90}$ &  $\mathbf{0.91}$\\
                                     & $\%outlier$ &	$\mathbf{3.57}$ & $\mathbf{3.57}$ & $\mathbf{3.57}$ & $7.14$ & $3.57$ & $5.36$ & $8.93$ & $1.79$ & $\mathbf{4.69}$\\
\hline
\multirow{2}{3.1em}{\textit{BaseDA}}  &	$DC$ (P50)    &	$0.94$ & $\mathbf{0.90}$ & $\mathbf{0.96}$ & $0.95$ & $\mathbf{0.91}$ & $\mathbf{0.93}$ & $\mathbf{0.93}$ & $\mathbf{0.91}$ & $\mathbf{0.93}$\\
                                     &	$DC$ (P10)    &	$0.92$ & $\mathbf{0.87}$ & $0.93$ & $\mathbf{0.94}$ & $\mathbf{0.90}$ & $\mathbf{0.92}$ & $\mathbf{0.92}$ & $\mathbf{0.90}$ & $\mathbf{0.91}$\\
                                     & $\%outlier$ &	$5.36$ & $\mathbf{3.57}$ & $7.14$ & $\mathbf{1.79}$ & $1.79$ & $8.93$ & $7.14$ & $\mathbf{0.00}$ & $4.85$\\
\hline
\multirow{2}{3.1em}{\textit{BaseMS}}   & $DC$ (P50)    &	$0.94$ & $0.87$ & $0.95$ & $0.95$ & $0.89$ & $0.91$ & $0.92$ & $0.89$ & $0.92$\\
                                     &	$DC$ (P10)    &	$0.92$ & $0.84$ & $0.92$ & $0.93$ & $0.87$ & $0.89$ & $0.90$ & $0.87$ & $0.89$\\
                                     & $\%outlier$ &	$\mathbf{3.57}$ & $7.14$ & $5.36$ & $8.93$ & $\mathbf{0.00}$ & $\mathbf{5.36}$ & $\mathbf{5.36}$ & $7.14$ & $5.06$\\
\hline
\end{tabular}
\label{table:oasis-test}
\end{table}

\begin{table*}[t!]
\centering
\scriptsize
\caption{Summary of segmentation results on the MICCAI 2012 and MS Dataset. DC: Dice scores. Best results are highlighted.}
\setlength{\tabcolsep}{4pt} 
\renewcommand{\arraystretch}{1.1} 
\begin{tabular}{l|c|cccccccc | c || cccccccc | c}
\hline
Model                               &	Metrics     &	\multicolumn{9}{l}{MICCAI 2012 (Manual delineations)} 						  & \multicolumn{9}{l}{MS Dataset -- test (Automated Delineations)} \\

        &       									&		WM &	GM          & LV   	 &	Thal  &	HC    &	CdN     & Pu    & GP      &  All & WM   &	GM  & LV   &	Thal &	HC   &	CdN & Pu & GP & All\\
\hline\hline
\multirow{2}{3.1em}{\textit{Base}}     &	$DC$ (P50)    &	$0.89$ & $\mathbf{0.87}$ & $0.88$ & $0.85$ & $0.80$ & $0.80$ & $\mathbf{0.83}$ & $\mathbf{0.78}$ & $\mathbf{0.84}$ & $0.90$ & $0.87$ & $0.93$ & $0.92$ & $0.89$ & $0.91$ & $0.90$ & $0.87$ & $0.90$\\
                                     &	$DC$ (P10)    &	$0.85$ & $0.82$ & $0.82$ & $0.83$ & $0.77$ & $0.74$ & $\mathbf{0.80}$ & $0.73$ & $0.80$ & $0.86$ & $0.83$ & $0.87$ & $0.87$ & $0.78$ & $0.86$ & $0.82$ & $0.75$ & $0.83$\\
                                     & $\% outlier$ &	$\mathbf{3.33}$ & $3.33$ & $3.33$ & $6.67$ & $0.00$ & $\mathbf{0.00}$ & $6.67$ & $10.0$ &  $4.17$ & $8.99$ & $7.87$ & $\mathbf{2.25}$ & $6.74$ & $13.5$ & $4.49$ & $10.1$ & $10.1$ & $8.01$\\
\hline
\multirow{2}{3.1em}{\textit{BaseDA}}  &	$DC$ (P50)    &	$0.89$ & $0.86$ & $0.88$ & $\mathbf{0.86}$ & $\mathbf{0.81}$ & $0.81$ & $0.81$ & $\mathbf{0.78}$ & $\mathbf{0.84}$ &  $0.91$ & $0.88$ & $0.94$ & $0.93$ & $0.89$ & $\mathbf{0.92}$ & $0.91$ & $\mathbf{0.90}$ & $0.91$\\
                                     &	$DC$ (P10)    &	$0.87$ & $\mathbf{0.83}$ & $0.82$ & $\mathbf{0.84}$ & $\mathbf{0.78}$ & $0.73$ & $0.79$ & $\mathbf{0.74}$ & $0.80$ & $0.89$ & $0.85$ & $0.89$ & $0.92$ & $\mathbf{0.88}$ & $0.89$ & $0.87$ & $0.86$ & $0.88$\\
                                     & $\% outlier$ &	$\mathbf{3.33}$ & $3.33$ & $\mathbf{0.00}$ & $\mathbf{0.00}$ & $0.00$ & $3.33$ & $\mathbf{3.33}$ & $\mathbf{6.67}$ & $\mathbf{2.50}$ & $\mathbf{3.37}$ & $\mathbf{0.00}$ & $5.62$ & $\mathbf{2.25}$ & $\mathbf{1.12}$ & $4.49$ & $6.74$ & $\mathbf{4.49}$ & $\mathbf{3.51}$\\
\hline
\multirow{2}{3.1em}{\textit{BaseMS}}  &	$DC$ (P50)    &	$\mathbf{0.90}$ & $0.84$ & $0.88$ & $0.85$ & $0.80$ & $\mathbf{0.82}$ & $0.82$ & $0.76$ & $0.83$ & $\mathbf{0.94}$ & $\mathbf{0.90}$ & $\mathbf{0.95}$ & $\mathbf{0.94}$ & $\mathbf{0.90}$ & $\mathbf{0.92}$ & $\mathbf{0.92}$ & $\mathbf{0.90}$ & $\mathbf{0.92}$\\
                                     &	$DC$ (P10)    &	$\mathbf{0.88}$ & $0.80$ & $0.82$ & $0.83$ & $0.77$ & $\mathbf{0.77}$ & $0.79$ & $0.73$ & $0.80$ & $\mathbf{0.92}$ & $\mathbf{0.87}$ & $\mathbf{0.91}$ & $\mathbf{0.93}$ & $\mathbf{0.88}$ & $\mathbf{0.90}$ & $\mathbf{0.91}$ & $\mathbf{0.87}$ & $\mathbf{0.90}$\\
                                     & $\% outlier$ &	$10.0$ & $\mathbf{0.00}$ & $3.33$ & $\mathbf{0.00}$ & $0.00$ & $6.67$ & $6.67$ & $\mathbf{6.67}$ & $4.17$ & $\mathbf{3.37}$ & $\mathbf{0.00}$ & $4.49$ & $6.74$ & $7.87$ & $\mathbf{3.37}$ & $\mathbf{4.49}$ & $7.87$ & $4.78$\\
\hline
\end{tabular}
\label{table:miccai-ms}
\end{table*}

\begin{figure*}[t]
    \centering
    \includegraphics[width=0.95\textwidth]{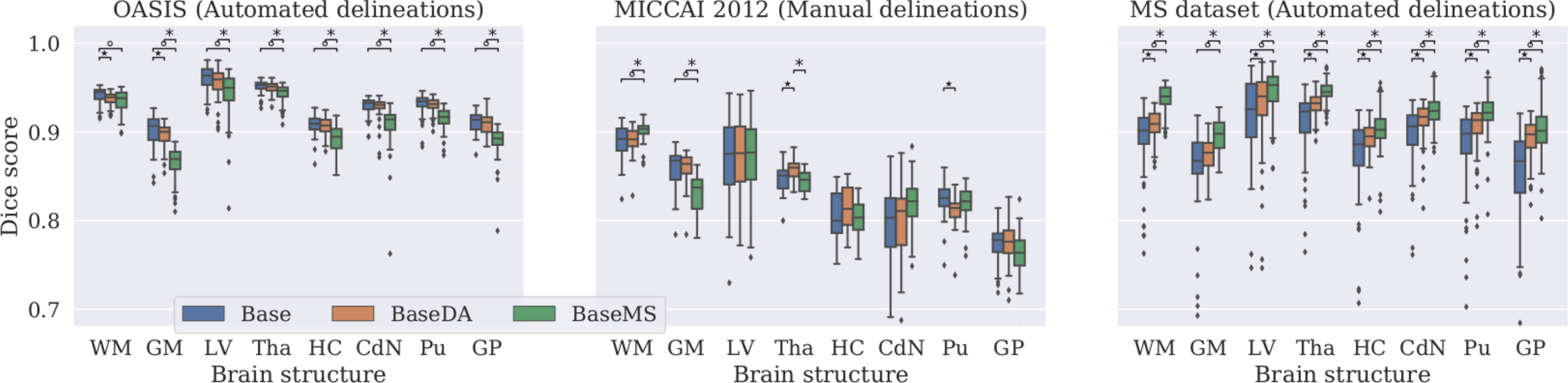}
    \caption{Dice scores on all the test sets for the three models. Statistically different results (Wilcoxon, $p < 0.05$) are indicated: ($\star$: \textit{Base} and \textit{BaseDA} are different; $\circ$: \textit{Base} and \textit{BaseMS} are different; $*$: \textit{BaseDA} and \textit{BaseMS} are different).} 
    \label{fig:boxplots_results}
\end{figure*}

\noindent \textbf{MICCAI 2012}
On the manual labels, the different models reach comparable performance, with very few statistical differences (see Fig. \ref{fig:boxplots_results} and Table \ref{table:miccai-ms}). 
Variances are not statistically different for any tissue type, except for WM between \textit{Base} and \textit{BaseMS}, where the latter has lower variance (Levene: $p=0.04$).
$Se$ and $Pr$ vales are also comparable for all models, with   mean} $\bar{Se}\approx0.84, \bar{Pr}\approx0.85$.
Here all models achieve lower performance, which is expected since they were trained on automated delineations.

\noindent \textbf{MS Dataset}
\textit{BaseDA} outperforms \textit{Base} for all structures (Wilcoxon: $p\ll0.05$ for all except HC: $p<0.05$,  GM: $p>0.05$). $\% outliers$ is consistently larger for the \textit{Base} model and the variances for this model are larger for all the tissue types (Levene: $p\ll0.05$ for all except CdN: $p<0.05$ and LV: $p>0.05$).
As expected, \textit{BaseMS} is still better at generalizing to this type of multi-center data. As illustrated in Fig. \ref{fig:boxplots_results}, the mean $DC$ is consistently larger than for the other two models (Wilcoxon, $p\ll0.05$). \textit{Base} is always worse in terms of variance, but \textit{BaseDA} approximates the variability of the \textit{BaseMS} models, except for the case of LV and HC (Levene: $p<0.05$).
$Se$ values are also lower in the \textit{Base} model (min: $Se_{GP}=0.82,$     mean: $\bar{Se}=0.88$), while in the other models these values remain comparable to the values reported for the OASIS test set.

It is important to keep in mind that the MS Dataset contains pathological images which are not present in OASIS. \textit{BaseMS} has been exposed to many more types of images, with some patients possibly presenting a small number of lesions. However, the contrary is not true, given that OASIS only contains images from healthy patients. At best, the networks trained on this data were exposed to a few lesions present in the older patient’s scans. It is thus not possible to guarantee that the differences in performance between \textit{BaseMS} and the other models   on a pathological dataset are caused only by scanner variability.

\section{Conclusions and future work}

We present a novel intensity-based data augmentation strategy with the goal of generalizing models trained on homogeneous datasets to multi-scanner and multi-center data. This is a fast and simple method which can be added to the training pipeline to generate images on-the-fly. The method is very general and can potentially be used in other MRI applications. 

We restricted the data augmentation procedures to a minimum so that we could observe the effect of adding the instensity transformation alone. Additionally, since the images were registered to MNI space we decided not to include any geometric transformations. 
It is worth noting that the DA algorithm still works well if the images are in native space. Registration was performed as a way to simplify the learning of the network, since we were interested in comparing the effect of the augmentation step in a simplified setting. 

There are a few limitations to the present work. Namely, due to scarcity of manual delineations, the methods were trained on automated segmentations. This is not ideal, because our method is likely to inherit any bias or know problems that might exist in the ground truth. However, given that we are especially interested in the effect of the augmentation we can still make a fair comparison between the approaches.
Additionally, the presence of pathology in the MS Dataset introduces an extra source of variability. In the future we intend to explore adding one more component to the mixture to account for intensity changes due to disease. Even when training on OASIS, it is likely that older subjects present white matter abnormalities that resemble MS lesions. This remains outside of the scope of the present work. Finally, all images were bias-field corrected as a pre-processing step. We would like to note that the GMM-based method still works on images with bias field, but we did not explore how the results are influenced if at test time we do not correct for the bias field. In the future we plan to explore how adding a bias-field augmentation procedure after the intensity augmentation would affect final results.

Another open question is whether adding this method to an already heterogeneous dataset would improve the performance, but additional experiments are necessary to verify this. 
Future work includes expanding the method to allow alterations to the shape of the distributions of the different components, leaving the normality assumption.

\section{Compliance with Ethical Standards}
\label{sec:acknowledgments}
This research study was conducted retrospectively using human subject data partly made available in open access by OASIS \cite{Marcus2007} and and manual labelings by Neuromorphometrics, Inc. under academic subscription  \cite{MICCAI2012}.
Ethical approval was not required as confirmed by the license attached with the open access data. 
The MS dataset is a subset of data processed with ico\textbf{brain ms} in clinical practice, for which subjects had agreed to allow ico\textbf{metrix} to use an anonymised version of the already analysed MR images for research purposes. 

\section{Acknowledgments}

This work was supported by the European Union’s Horizon 2020 research and innovation program under the Marie Sklodowska-Curie grant agreements No 765148 and No 764513 and by the NIH NINDS grant No R01NS112161. 

\small
\bibliographystyle{IEEEbib}
\bibliography{refs}

\end{document}